\documentclass[10 pt, conference]{IEEEtran}

\usepackage[ruled]{algorithm2e}

\usepackage{pdfpages}
\usepackage{subfigure,graphicx}
\usepackage{amsmath,epsfig}
\usepackage{color}
\usepackage{algpseudocode}
\usepackage{epstopdf}

\newtheorem{myDef}{\textbf{Definition}}
\newtheorem{myTheo}{\textbf{Theorem}}
\newtheorem{myPro}{\textbf{Proposition}}

\usepackage{cite}
\usepackage{amsmath}

\vspace{-4mm}

\begin{document}

\title{When Data Sponsoring Meets Edge Caching: \\ A Game-Theoretic Analysis}

\IEEEoverridecommandlockouts

\author{
\IEEEauthorblockN{Haitian Pang\IEEEauthorrefmark{1},
Lin Gao\IEEEauthorrefmark{2},
Qinghua Ding\IEEEauthorrefmark{1},
Lifeng Sun\IEEEauthorrefmark{1}}
\IEEEauthorblockA{\IEEEauthorrefmark{1}
Tsinghua National Laboratory for Information Science and Technology, and
\\
Department of Computer Science and Technology, Tsinghua University, Beijing, China}
\IEEEauthorblockA{\IEEEauthorrefmark{2}
School of Electronic and Information Engineering, Harbin Institute of Technology, Shenzhen, China}
\IEEEauthorblockA{Email: \{pht14@mails., dqh14@mails., sunlf@\}tsinghua.edu.cn, gaol@hit.edu.cn}
\thanks{This work is supported by the National Natural Science Foundation of China (Grant No. 61771162, 61472204, and 61521002), Shenzhen Science and Technology Plan (Grant No. CXZZ20150323151850088), and Beijing Key Laboratory of Networked Multimedia (Grant No. Z161100005016051).}
\vspace{-8mm}
}

\maketitle

\begin{abstract}
\emph{Data sponsoring} is a widely-used incentive method in today's cellular networks, where video content providers (CPs) cover part or all of the cellular data cost for mobile users so as to attract more video users and increase data traffic.
In the forthcoming 5G cellular networks, \emph{edge caching} is emerging as a promising technique to deliver videos with lower cost and higher quality. The key idea is to cache video contents on edge networks (e.g., femtocells and WiFi access points) in advance and deliver the cached contents to local video users directly (without involving cellular data cost for users).
In this work, we aim to study how the edge caching will affect the CP's data sponsoring strategy as well as the users' behaviors and the data market.
Specifically, we consider a single CP who offers both the edge caching service and the data sponsoring service to a set of heterogeneous mobile video users (with different mobility and video request patterns).
We formulate the interactions of the CP and the users as a two-stage Stackelberg game, where the CP (leader) determines the budgets (efforts) for both services in Stage I, and the users (followers) decide whether and which service(s) they would like to subscribe to.
We analyze the sub-game perfect equilibrium (SPE) of the proposed game systematically.
Our analysis and experimental results show that by introducing the edge caching, the CP can increase his revenue by $105\%$.
\end{abstract}


\section{Introduction}
\label{sec:intro}

\subsection{Background and Motivations}

Nowadays, we are witnessing the explosive growth of global mobile data traffic.
According to Cisco \cite{cisco}, mobile video traffic accounts for a majority of the total mobile traffic (e.g., $60\%$ in 2016).
Due to the fast increase of video traffic, the increased data cost is becoming one of the major concerns for mobile users to watch videos \cite{add-1}.
This brings additional challenge (and also huge opportunity) for video content providers (CPs), as they need to consider not only the quality improvement of their offered video services as before, but also the cost reduction for the users who request their services (to attract more video users).
Due to the intensive competition of CPs, the second issue (i.e., reducing user cost) is becoming increasingly important today \cite{add-2,add-4}.

One effective way to reduce user cost is the so-called \emph{data sponsoring} \cite{sdp1,sdp2,sdp3}, which has been employed by many CPs worldwide.
The key idea is to allow CPs to \emph{subsidize} the users' cost of mobile video data, hence attract more mobile video users and traffic.
Data sponsoring creates a win-win situation for mobile users and CPs, that is,  mobile users benefit from the free access of video contents, and CPs benefit from the increased video users and traffic (through, for example, selling more built-in advertisements).
Thus, we refer to such a data sponsoring scheme as ``\emph{cellular data sponsoring}''.
As a real-world example, AT\&T announced its sponsored data program in January 2014 \cite{att}.

In the forthcoming 5G cellular networks, \emph{edge caching} is emerging as a promising technique to deliver videos with lower cost and higher quality \cite{sam,add-5}.
The key idea is to cache the popular video contents on edge networks (e.g., femtocell base stations and WiFi access points) in advance and deliver the cached contents to local video users directly via device-to-device connections (e.g., WiFi direct).
Obviously, with edge caching, mobile users can obtain video contents without incurring the cellular data cost.
In this sense, edge caching can be viewed as a new sponsorship scheme for mobile users.
We refer to such a new sponsoring scheme as ``\emph{edge cache sponsoring}''.
As an example, Xunlei \cite{thunder}, one of the largest content delivery networks (CDNs) in China, has deployed WiFi APs with large storage capacity to deliver video contents for mobile users.

In this work, we will study the mobile video data market with both sponsoring schemes (as in \cite{joint}).
We aim to understand how the newly introduced edge caching will affect the traditional cellular data sponsoring, and how the co-existence of edge caching and cellular data sponsoring will change the user behavior and the whole data market.

\subsection{Solution and Contributions}

To concentrate on the mutual interaction of edge caching and cellular data sponsoring, we consider a simple model with a \emph{single CP}, who offers both the edge cache sponsoring and the cellular data sponsoring to mobile users.
As in the existing literature \cite{thunder,joint, ec2}, we assume that the cellular network is available in the whole area, while the edge network is only available in part of the area (e.g., hotspots) due to the limited distance of device-to-device transmission.

By covering the data cost for users with either edge cache sponsoring or cellular data sponsoring, the CP can attract more video users and traffic, and hence achieve certain  \emph{revenue gain} (e.g., via build-in advertisements).
When providing sponsoring for users, the CP needs some \emph{budget} for covering the cellular data cost (in cellular data sponsoring) or caching the video contents on edge network (in edge cache sponsoring), hence lead to certain \emph{revenue loss}.
Note that a higher budget for cellular data sponsoring implies that the CP will sponsor more video contents (for those users subscribing to the cellular data sponsoring), and a higher CP budget for edge cache sponsoring implies that the CP will cache more video contents on edge network (for those users subscribing to the edge cache sponsoring).
Clearly, a higher budget (effort) for a particular sponsoring scheme can attract more users to subscribe to it (hence bring more revenue gain for the CP), but will also introduce more loss to the CP to offer the sponsoring.
Thus, the CP needs to determine the budget for each sponsoring scheme carefully to balance the revenue gain and loss.

Given the budgets that the CP offers for both sponsoring schemes, mobile users will decide whether and which sponsoring scheme(s) they are going to choose, which lead to the following four different memberships:
\begin{itemize}
\item \textbf{NoSp}: Choosing neither sponsoring scheme;
\item \textbf{CellSp}: Choosing cellular data sponsoring;
\item \textbf{EdgeSp}: Choosing edge cache sponsoring;
\item \textbf{HybridSp}: Choosing both sponsoring scheme.
\end{itemize}

More specifically, when a user chooses CellSp, his request will be served by the cellular network, and the cellular data cost will be covered by the CP with a given probability (depending on the CP's budget for   cellular data sponsoring).
When a user chooses EdgeSp, his request will be served by the edge network if the edge cache is available (i.e., the user is within the edge network and meanwhile the requested contents have been cached on the edge network).
When a user chooses HybridSp, his video request will be served by the edge network if the edge cache is available, or otherwise, by the cellular network and the cellular data cost will be covered by the CP with a given probability.

Game theory \cite{gametheory} has been widely used in wireless networks (e.g., \cite{gao-1,gao-2,gao-3,gao-4,gao-5,gao-6}) for modeling and analyzing the competitive and cooperative interactions among different network entities.
We formulate the interactions of the CP and the users as a two-stage Stackelberg game \cite{gametheory}, where the CP acts as game \emph{leader} determining the budgets for both sponsoring schemes in Stage I, and the users act as game \emph{followers} deciding which membership they would like to choose. 
We analyze the sub-game perfect equilibrium (SPE) of the proposed game systematically.
In summary, the key contributions of this works are given below.
\begin{itemize}
\item \emph{Novel Model}:
    This work analyzes the scenario where both cellular data sponsoring and edge caching are provided to users simultaneously.
    Our model captures many important features of practical systems, such as the content popularity and user heterogeneity.

\item \emph{Game-Theoretic Analysis}:
    We formulate the problem as a Stackelberg game and provide a comprehensive game theoretic analysis.
    We prove the uniqueness and existence of equilibrium in Stage II, and employ numerical method to obtain equilibrium in Stage I.

\item \emph{Experiments and Insights}:
    We conduct extensive experiments to evaluate the system performance.
    The experiment results show that by introducing the edge caching, the CP can increase his revenue by $105\%$. Users with high probability to meet an edge device prefer the edge cache sponsoring. 
\end{itemize}

The rest of this paper is organized as follows.
In Section II, we present the system model.
In Section III, we provide the game formulation.
In Section IV, we provide the game equilibrium analysis.
We provide simulation results in Section V, and conclude in Section VI.

\section{System Model}

\subsection{Network Model}
We consider a single CP providing video service to a set $\mathcal{U} = \{1,2,...,U\}$ of mobile users. The CP can deliver video contents to mobile users in two different ways:
\begin{itemize}
  \item \emph{Cellular Direct Delivery}: The video content (located in the remote server) will be delivered to the user through the cellular link directly;
  \item \emph{Edge Cache Delivery}: The video content is cached in the edge network in advance, and will be delivered to the (nearby) users through the local link.
\end{itemize}
Each user moves and requests video contents randomly according to his own pattern. Let $\mathcal{S} = \{1,2,...,S\}$ denote the set of all video contents.

To enable the edge cache delivery, the CP needs to cache the video contents in the edge network devices (at the corresponding locations) in advance. Thus, a video request can be delivered via the edge network only when the user is within the edge network and the requested content is cached.

\subsection{CP Model}\label{sec:cpmodel}
The CP's goal is to maximize his total sponsoring revenue, which depends on the revenue loss induced by the budget and the obtained revenue gain from the sponsoring. The CP optimizes his revenue via managing the budgets in sponsoring schemes. When a user's request is sponsored through cellular delivery, the CP will cover the cellular data cost for the user. We denote the cellular data cost of one request for the CP as $h_1$. The total cost spent on cellular sponsoring is proportional to the total data volume of cellular sponsored requests. We denote the cellular sponsor effort as $\alpha_1$, hence the total cost in cellular sponsoring is $h_1 \alpha_1$. The cost in edge cache delivery is the storage cost of cached contents. We denote $h_2$ as the edge cache cost for one content per time period. We denote the edge cache sponsor effort as $\alpha_2$, hence the total cost in edge cache sponsoring is $h_2 \alpha_2$.

Moreover, if a content is sponsored (either via cellular sponsoring or edge cache sponsoring), it will be delivered to the user with certain attachment (e.g., build-in advertisements) called \emph{value-added content}, hence can bring revenue gain for the CP. We define the average revenue gain of a request for the CP as $u$.

\subsection{User Model}
Considering the heterogeneity of users in mobility pattern and service request pattern, we define the probability that a user is covered by the edge network as $r_u \in [0,1]$, and the probability to request a content as $f_u \in [0,1]$. In reality, to allow the user to estimate his mobility probability and request probability, we assume that the CP will announce the related information at the beginning. Specifically, the CP will announce the locations to place the edge caching devices, and the content caching rule, e.g., the CP will cache the latest popular TV series and talk shows. After the announcement made by the CP, the users can estimate the above probabilities based on their mobility and request patterns.

In this work, we focus on the \emph{symmetric equilibrium} where users with the same type $(f, r)$ will always make the same membership choice. Moreover, we focus on the user \emph{pure-strategy behavior} where each user will choose a specific membership under a given network situation. Namely, a user makes his membership choice among \textbf{NoSp}, \textbf{CellSp}, \textbf{EdgeSp}, and \textbf{HybridSp}. For notation convenience, we denote the membership as $m \in \{N, C, E, H\}$, correspondingly.

We denote $\Theta_N$, $\Theta_C$, $\Theta_E$, and $\Theta_H$ as the sets of users choosing \textbf{NoSp}, \textbf{CellSp}, \textbf{EdgeSp}, and \textbf{HybridSp}. Hence, the percentages of $C$, $E$, and $H$ users are: $\mu_C=\frac{|\Theta_C|}{U}$, $\mu_E=\frac{|\Theta_E|}{U}$, $\mu_H=\frac{|\Theta_H|}{U}$.

 The \emph{payoff} of each user is the achieved benefit minus the incurred cost. For convenience, we denote the payoff of a type-$(f, r)$ user under membership $m$ as $V_{(f, r)}(m)$. The objective of the user is to choose the proper membership to maximize his payoff.

 \subsubsection{\textbf{NoSp}}
 A user choosing neither sponsoring scheme cannot connect to any sponsoring network. As a user choosing \textbf{NoSp} has no cost or revenue in the sponsoring network, we set the user payoff as zero: \begin{equation}V_{(f, r)}(N)=0.\end{equation}
	
 \subsubsection{\textbf{CellSp}}
 A user choosing the cellular data sponsoring can connect to the cellular sponsoring network at any locations for any contents. Due to the cellular budget constraint of the CP, the probability that a content request is sponsored via cellular network is $P$ (derived in Section \ref{sec:variables}). Then, the expected payoff (per time slot) of a type-$(f, r)$ user is
\begin{equation}V_{(f, r)}(C)=f\cdot P\cdot (v-c_1)-\phi_1,\end{equation}
where $v$ is the average benefit of a user request, $c_1$ is the average energy cost of a request, and $\phi_1$ is the cost of joining the \textbf{CellSp} (e.g., time-average energy cost). Note that we assume that the user benefit and the user cost are of the same unit.
	
 \subsubsection{\textbf{EdgeSp}}
A cache sponsored user can only connect to the sponsor network only when he is within the edge network and requests the cached video content \cite{mobilitypred}. If a user is sponsored from the edge network, the video is delivered from the edge device to the user via local network connection. Let $\rho$ denote the probability of a user requesting a cached content in a time slot (derived in Section \ref{sec:variables}). Then, the expected payoff (per time slot) of a type-$(f, r)$ user is:
 \begin{equation}V_{(f, r)}(E)=r\cdot f\cdot\rho \cdot (v-c_2)-\phi_2,\end{equation}
where $r$ is the user mobility factor, $v$ is the average value of a request for the user, $c_2$ is the average energy cost of a request via the edge network, and $\phi_2$ is the cost of joining the \textbf{EdgeSp}.

 \subsubsection{\textbf{HybridSp}}
A user can choose to join both the two sponsoring schemes. When he is within the cache region and hits the cached content, he can employ the edge cache delivery. Otherwise, he will compete for the cellular sponsoring. Hence, the expected payoff (per time slot) of a type-$(f, r)$ user is:
\begin{equation}V_{(f, r)}(H)=(f-r\cdot f\cdot\rho)\cdot P\cdot (v-c_1)+r\cdot f\cdot\rho \cdot (v-c_2)-\phi_1-\phi_2,\end{equation}

\section{Game Formulation}

Based on the system model above, we formulate the interaction between the CP and mobile users as a two-stage Stackelberg game.

In Stage I, the CP decides the sponsor budgets (efforts) of cellular and edge cache sponsoring: $\alpha_1$, $\alpha_2$, respectively, to maximize the expected profit. We employ the widely-adopted contracted sponsoring for cellular sponsoring like in \cite{pricing}. The CP will pay a fixed fee $\alpha_1 h_1$ at first. Thus the expected revenues (per time slot) of the cellular and edge cache sponsoring are
\begin{equation}U_C(\alpha_1, \alpha_2)=u\cdot P\cdot N_C  - \alpha_1 h_1,\end{equation}
\begin{equation}U_E(\alpha_1, \alpha_2)= u\cdot N_E - \alpha_2 h_2,\end{equation}
where $u$ is the CP revenue defined in Section \ref{sec:cpmodel}, $N_C$ and $N_E$ denote the expected requests in cellular sponsoring and edge cache sponsoring, respectively, $P$  denotes the probability for users to get sponsoring in the cellular network. Hence we can define the sum CP revenue as
\begin{equation}U(\alpha_1, \alpha_2)=U_C(\alpha_1, \alpha_2)+U_E(\alpha_1, \alpha_2),\end{equation}
Keep in mind that $N_C$, $N_E$, and $P$ are all functions of $(\alpha_1, \alpha_2)$ derived in Stage II.

In Stage II, users determine their memberships (i.e., \textbf{NoSp}, \textbf{CellSp}, \textbf{EdgeSp}, or \textbf{HybridSp}), given the budgets $\alpha_1$ and $\alpha_2$ announced by the CP in Stage I. Note that users' decisions are coupled. For example, with more users choosing \textbf{CellSp}, the user payoff of cellular sponsor will decrease (due to a smaller sponsor probability $P$), resulting in the payoff gain in edge cache sponsor. Similarly, with more users choosing \textbf{EdgeSp}, the payoff of the cellular sponsoring will increase (due to a larger sponsor probability $P$).

Next, we will study the Subgame Perfect Equilibrium (SPE) of the proposed two-stage Stackelberg game. We show the definition of the SPE in Definition 1.

\begin{myDef}
	A strategy profile $\{(\alpha_1^*, \alpha_2^*),(m^*(f, r)$, $\forall r\in [0,1], f\in [0,1]\}$, where $(\alpha_1^*, \alpha_2^*)$ is the CP's strategy in Stage I and $m^*(f, r)$ is a type-$(f, r)$ user's strategy in Stage II, is an SPE if and only if:
\begin{equation}
\left\{
\begin{aligned}
Stage~II: & V_{(f, r)}(m^*(f, r)) \ge V_{(f, r)}(m), \\
& \forall r\in [0,1], f\in [0,1], m \in\{ C, E, H, N\};  \\
Stage~I: & U(\alpha_1^*,\alpha_2^*) \ge U(\alpha_1^\prime,\alpha_2^\prime),\\
& \forall \alpha_1 \in [\alpha_{\min},\alpha_{\max}], \alpha_2 \in [\alpha_{\min},\alpha_{\max}], \\
\end{aligned}
\right.
\end{equation}
 where $\alpha_1^\prime$, $\alpha_2^\prime$ are all the other budget selections, $\alpha_{\min}$, $\alpha_{\max}$ are the lower bound and upper bound of the sponsor budget, respectively. For fairness, we set the budget bounds of cellular sponsoring and edge cache sponsoring as the same.
\end{myDef}

We will derive the SPE by backward induction. Namely, we first study the user membership selection game in Stage II, and derive the users' equilibrium decisions. Then we characterize the optimal budgets of the CP to maximize his profit in Stage I.

\subsection{Stage II: User Membership Selection}
As discussed previously, the sets of users choosing \textbf{NoSp}, \textbf{CellSp}, \textbf{EdgeSp}, and \textbf{HybridSp} are already in the system (i.e., $\Phi_N$, $\Phi_C$, $\Phi_E$, $\Phi_H$) and their corresponding percentages (i.e., $\mu_N$, $\mu_C$, $\mu_E$ and $\mu_H$) will affect the value of $P$, and further affect the user payoff and membership selection. Hence, we will first study what is the user's best membership decision under a particular membership distribution $\{\Phi_N, \Phi_C, \Phi_E, \Phi_H\}$. Then we will study how the user membership decision dynamically evolves over time, and what is the stable membership distribution (called \emph{membership selection equilibrium}).

Given the CP's strategy $(\alpha_1, \alpha_2)$, and under a particular initial membership distribution $\{\Phi_N, \Phi_C, \Phi_E, \Phi_H\}$, the payoff of a type-$(f, r)$ user is:
\begin{equation}\label{equa:userpayoff}
\begin{array}{l}
V_{(f, r)}(m)=
\left\{\begin{array}{ll}
0, & m=N,\\
(v-c_1)Pf-\phi_1, &m= C,\\
(v-c_2)\rho fr-\phi_2, & m=E,\\
 (v-c_2)\rho fr+(v-c_1)Pf-\\
 ~~~~(v-c_1)P\rho fr-\phi_1-\phi_2, & m=H.
\end{array}\right.
\end{array}
\end{equation}


A type-$(f, r)$ user will  choose (i)  \textbf{CellSp}, if and only if
\begin{equation}V_{(f, r)}(m=C)\ge \max\{0, V_{(f, r)}(E), V_{(f, r)}(H)\}, \end{equation}
(ii)   \textbf{EdgeSp}, if and only if
\begin{equation}V_{(f, r)}(m=E)\ge \max\{0, V_{(f, r)}(C), V_{(f, r)}(H)\}, \end{equation}
(iii)   \textbf{HybridSp}, if and only if
\begin{equation}V_{(f, r)}(m=H)\ge \max\{0, V_{(f, r)}(C), V_{(f, r)}(E)\}, \end{equation} 
and (iv) \textbf{HybridSp} if  the above conditions are not satisfied. 
	
\subsection{Stage I: CP Budget Selection}
	Given any CP's strategy $(\alpha_1, \alpha_2)$, the cellular sponsor probability $P^*(\alpha_1, \alpha_2)$, the number of user requests for cellular sponsor and edge cache sponsor $N^*_C(\alpha_1, \alpha_2)$ and $N^*_E(\alpha_1, \alpha_2)$ are achieved under the NE. Hence, we can formulate the CP payoff maximization problem as
	\begin{equation}
	\begin{aligned}
	\max_{\alpha_1, \alpha_2}X^*_C(\alpha_1, \alpha_2)+X^*_L(\alpha_1, \alpha_2)-\alpha_1 h_1-\alpha_2 h_2, \\
	\end{aligned}
	\end{equation}
	where $X^*_C(\alpha_1, \alpha_2)=u N^*_C(\alpha_1, \alpha_2)P^*(\alpha_1, \alpha_2)$
	and \\$X^*_L(\alpha_1, \alpha_2)= u N^*_E(\alpha_1, \alpha_2)$.

\subsection{Derivation of Important Variables}\label{sec:variables}
	Before studying the game equilibrium, we first derive the values of $\rho$, $N_C$, $N_E$, and $P$ analytically.
	
\subsubsection{\textbf{Calculation of $\rho$} (Cached Video Content Probability)}
 	We assume that the distribution of video requests follows Zipf distribution: $g(s) ,\;s\in \{1,2,...,S\}$. Given the budget $\alpha_2$ of edge cache sponsoring, we can derive the probability that a user requests a cached video content by:
	\begin{equation}\rho= \sum_1^{\alpha_2}g(s)\in [0,1].\end{equation}
Note that $\rho$ relies on the content number the CP decides to cache on each edge device.

\subsubsection{\textbf{Calculation of $N_C$ and $N_E$} (Expected Sponsoring Requests)}
	Given the users who choose \textbf{CellSp}, \textbf{EdgeSp}, and \textbf{HybridSp}, i.e.,  $\Theta_C$, $\Theta_E$, and $\Theta_H$, we can calculate the expected sponsoring request number:
	\begin{equation}
N_C=\sum_{u \in \Theta_C}f_u+\sum_{u \in \Theta_H}(f_u-r_u\cdot f_u\cdot\rho),\end{equation}
	\begin{equation}
N_E=\sum_{u \in \Theta_E\cup \Theta_H}r_u\cdot f_u\cdot\rho.~~~~~~~~~~~~~~~
\end{equation}

\subsubsection{\textbf{Calculation of $P$} (Sponsor Probabilities)}
	Given the budget $\alpha_1$ of cellular data sponsoring, the probability of users choosing \textbf{CellSp} or \textbf{HybridSp} can be computed as \begin{equation}P= \left\lceil \frac{\alpha_1}{N_C} \right\rceil^1,\end{equation}
where $\lceil x\rceil^1=\max(x,1)$.

\section{Game Equilibrium Analysis}
We now analyze the game equilibrium by using backward induction in this Section.
\subsection{User Selection Game in Stage II}
	
\subsubsection{Analysis of the NE}
 
Based on the user payoff formulated in (\ref{equa:userpayoff}), we analyze user selection distribution to get insight into the NE. We assume that $\phi_1 = \phi_2 = \phi$, as the energy spent to keep the communication with the CP is the same. Furthermore, we define $\delta_1=(v-c_1)P$ and $\delta_2=(v-c_2)\rho$ as the instant payoff of delivery networks. Hence, the user payoffs under different memberships can be reformulated as:
\begin{equation}
\begin{array}{l}
V_{(f,r)}(m)=
\left\{\begin{array}{ll}
0, & m=N,\\
\delta_1f-\phi, &m= C,\\
\delta_2fr-\phi, & m=E.\\
(\delta_2-\delta_1\rho)fr+\delta_1f-2\phi, &m=H
\end{array}\right.
\end{array}
\end{equation}

We can derive the user's selection policy as below. The mobile user will choose:

\begin{itemize}
\item m=N, if and only if $f<\frac{\phi}{\delta_1}$, $rf<\frac{\phi}{\delta_2}$.\footnote{The third condition that $(\delta_2-\delta_1\rho)rf+\delta_1f-2\phi<0$ is covered by the former two conditions, since we have $(\delta_2-\delta_1\rho)rf+\delta_1f-2\phi<(\delta_1f-\phi)+(\delta_2fr-\phi)<0$ }
\item m=C, if and only if $f>\frac{\phi}{\delta_1}$, $\delta_1-\delta_2r>0$ and $\phi-(\delta_2-\delta_1\rho)fr>0$.
\item m=E, if and only if $fr>\frac{\phi}{\delta_2}$, $\delta_2r-\delta_1>0$ and $\phi+\delta_1\rho fr-\delta_1f>0$.
\item m=H, if and only if $\phi-(\delta_2-\delta_1\rho)fr<0$, $\phi+\delta_1\rho fr-\delta_1f<0$ and $(\delta_2-\delta_1\rho)fr+\delta_1f-2\phi>0$.
\end{itemize}

We note that only when $\delta_i \ge \phi,\quad i=\{1,2\}$ exists, the sponsor via different networks benefit mobile users. Once $\delta_1, \delta_2$ are determined under the NE, we can derive the payoffs of a type-$(f, r)$ user under different memberships.

To facilitate the later analysis, we introduce the concept of \emph{indifferent point} (on the user type). 

\begin{myDef} An indifferent point is such a type-$(f, r)$ on which users will achieve the same payoff when selecting different memberships.\end{myDef}

\begin{figure}[t]
  \centering
     \includegraphics[width=.23\textwidth]{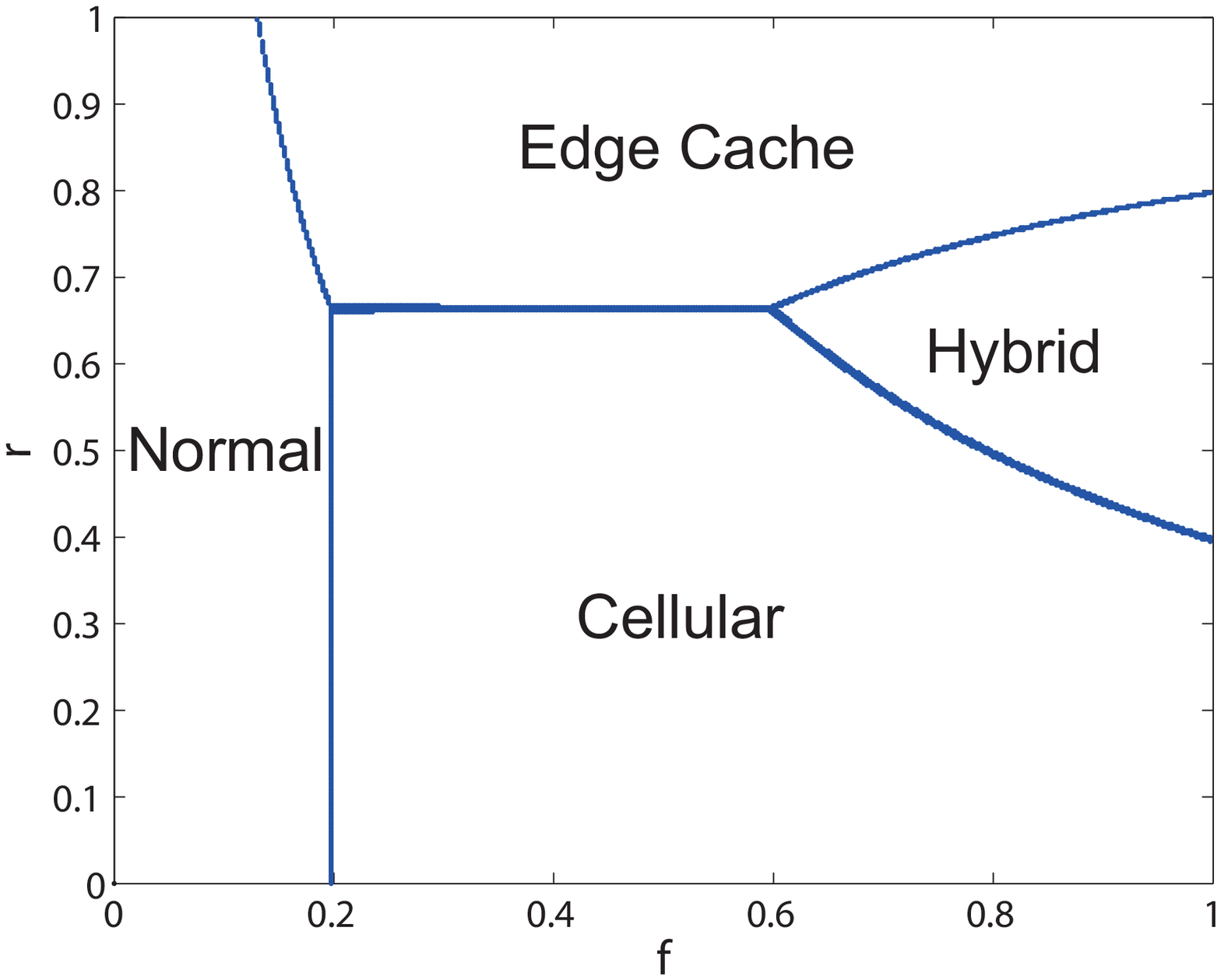}
   \includegraphics[width=.23\textwidth]{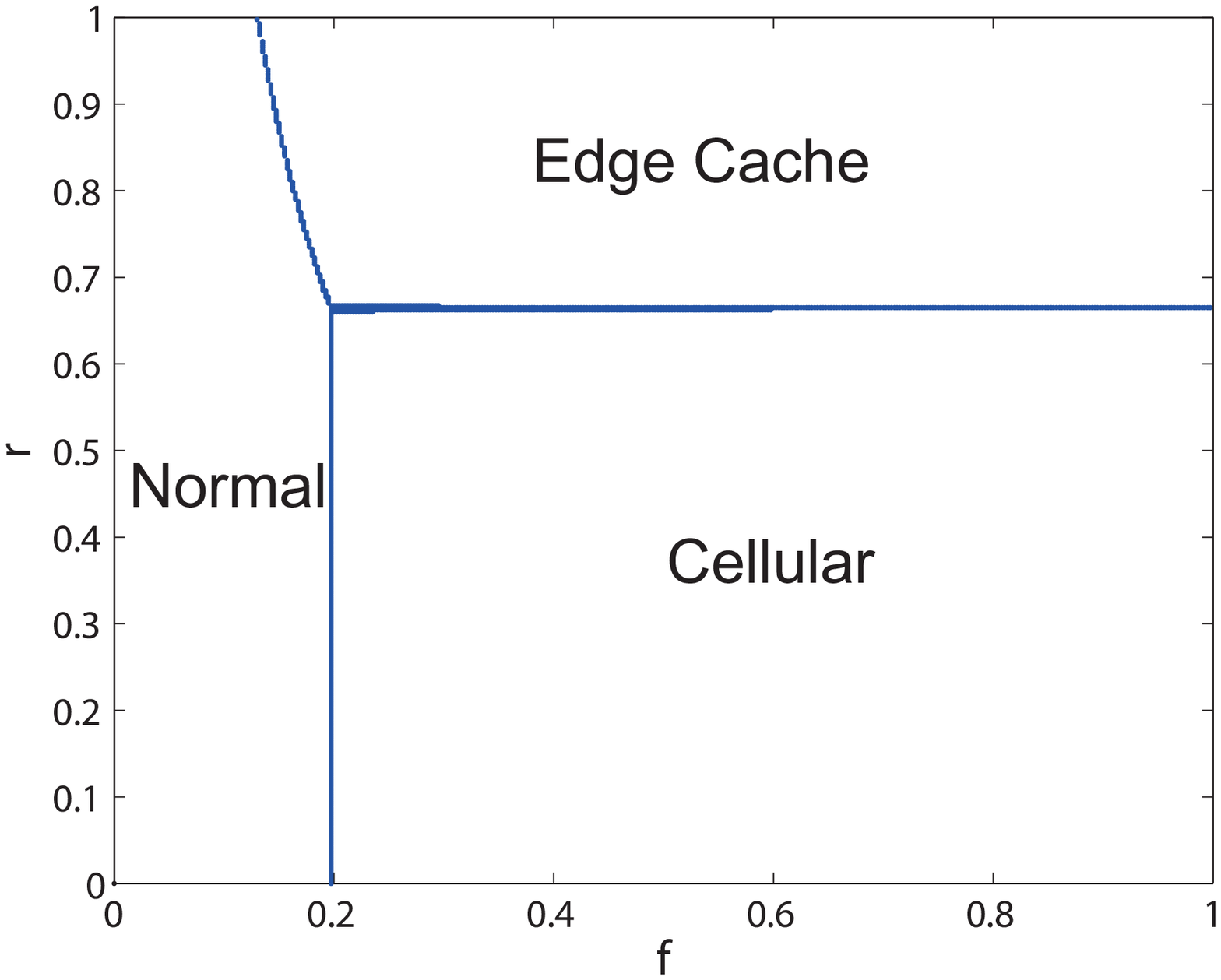}

	 \caption{User Divisions in NE with ``Hybrid'' and without ``Hybrid''.}\label{fig:udn}	
\vspace{-4mm}
\end{figure}

As shown in Fig.~\ref{fig:udn}, there are two kinds of indifferent points: (i) the type-$(f, r)$ where users will achieve the same payoff under \textbf{NoSp}, \textbf{CellSp}, and \textbf{EdgeSp}, and (2) the type-$(f, r)$ where users will achieve the same payoff under \textbf{HybridSp}, \textbf{CellSp}, and \textbf{EdgeSp}. Specifically,

\begin{enumerate}
\item \textbf{NoSp}, \textbf{CellSp}, and \textbf{EdgeSp} are the optimal membership selection for type-$(f, r)$ users where
\begin{equation}
\left\{\begin{array}{ll}
\delta_1f-\phi=0\\
\delta_2fr-\phi=0
\end{array}\right.
\end{equation}

\item \textbf{CellSp}, \textbf{EdgeSp}, and \textbf{HybridSp} are the optimal membership selection for type-$(f, r)$ users where
\begin{equation}
\left\{\begin{array}{ll}
\delta_1f-\phi=\delta_2fr-\phi\\
\phi+\delta_1\rho fr-\delta_1f=0
\end{array}\right.
\end{equation}
\end{enumerate}

We denote these two points as $N_1$ and $N_2$, respectively, where $N_1=(f^*,r^*)=(\frac{\phi}{\delta_1},\frac{\delta_1}{\delta_2})$ and $N_2=(\frac{f^*}{1-\rho r^*}, r^*)$. Next we discuss the existences of $N_1$ and $N_2$. 
\begin{enumerate}
\item Users will select \textbf{CellSp} and \textbf{EdgeSp}, if and only if $N_1$ exists and $N_2$ does not exist.
\item Users will select \textbf{CellSp}, \textbf{EdgeSp} and \textbf{HybridSp}, if and only if when both $N_1$ and $N_2$ exist.

\end{enumerate}

The user divisions in these two types of NE are illustrated in Fig.~\ref{fig:udn}. We can see that users with large $f$ and $r$ prefer edge cache sponsoring, and users with large $f$ and small $r$ prefer cellular sponsoring.

\subsubsection{Membership Distribution Dynamics}

Obviously, under the user best membership selection, the newly derived membership distributions may be different from the initial one. We denote the newly derived membership distribution by $\{\Theta_N^\prime, \Theta_C^\prime, \Theta_E^\prime, \Theta_H^\prime\}$, and the associated membership percentages as $\{\mu_N^\prime, \mu_C^\prime, \mu_E^\prime, \mu_H^\prime\}$.

The membership distribution continues evolving over time, until it reaches a stable distribution (called \emph{membership selection equilibrium}), where no user has the incentive to change his choice. Now we study the membership distribution dynamics, and characterize the membership selection equilibrium in Stage II.

Suppose that each user selects membership once in each time slot. Without loss of generality,  we consider the membership distribution change in a generic time slot $t$. 
 Let  $\{\Theta_N^t, \Theta_C^t, \Theta_E^t, \Theta_H^t\}$ denote the initial membership distribution at the beginning of slot $t$, and  $\{\Theta_N^{t+1}, \Theta_C^{t+1}, \Theta_E^{t+1}, \Theta_H^{t+1}\}$ denote the  newly derived membership distribution in time slot $t$ (after one round best response update). An \emph{equilibrium} is characterized by the following proposition.

  \begin{myPro}
  A membership distribution $\{\Theta_N^t, \Theta_C^t, \Theta_E^t$, $\Theta_H^t\}$ is a membership selection equilibrium
  if and only if $$\Theta_N^{t+1}=\Theta_N^t, \Theta_C^{t+1}=\Theta_C^t, \Theta_E^{t+1}=\Theta_E^t, \Theta_H^{t+1}=\Theta_H^t.$$
  \end{myPro}

This implies that if $\{\Theta_N^t, \Theta_C^t, \Theta_E^t, \Theta_H^t\}$ is a membership selection equilibrium, then we will have: $\Theta_m^\tau=\Theta_m^t, \forall m\in \{N,C,E,H\}$, for all $\tau>t$.

\begin{myTheo}\label{theo:s2ue}
There exists unique pure-strategy Nash Equilibrium (NE) in Stage II.
\end{myTheo}

\subsection{CP Budget Decision in Stage I}
Now we analyze the optimal decision of the CP in Stage I, given the NE of Stage II. Such analysis will lead to the SPE of the entire two-stage game.

We design a numerical method to solve the problem as follows.
We can check that the CP's optimization problem is non-convex. Hence, it is difficult to obtain the closed form solution of the optimal budgets $(\alpha_1^*, \alpha_2^*)$. Fortunately, the problem is a two variable optimization problem with box constraint sets, and can be solved using numerical methods. 
In this work, we solve the optimal budgets for the CP in a sequential manner: 
First, solve the optimal cellular budget 
$\alpha_1^*(\alpha_2)$ under any $\alpha_2$ and the optimal
$\alpha_2^*(\alpha_1)$ under any $\alpha_1$ through one-dimensional search; 
Then,
\begin{myPro}
The CP's best strategy $(\alpha_1^*, \alpha_2^*)$ must occur at an intersection point of $\alpha_1^*(\alpha_2)$ and $\alpha_2^*(\alpha_1)$.
\end{myPro}

\section{Numerical Results}

 \begin{figure*}[t]
  \centering

    \begin{minipage}[h]{.18\linewidth}
\includegraphics[width=0.9\textwidth]{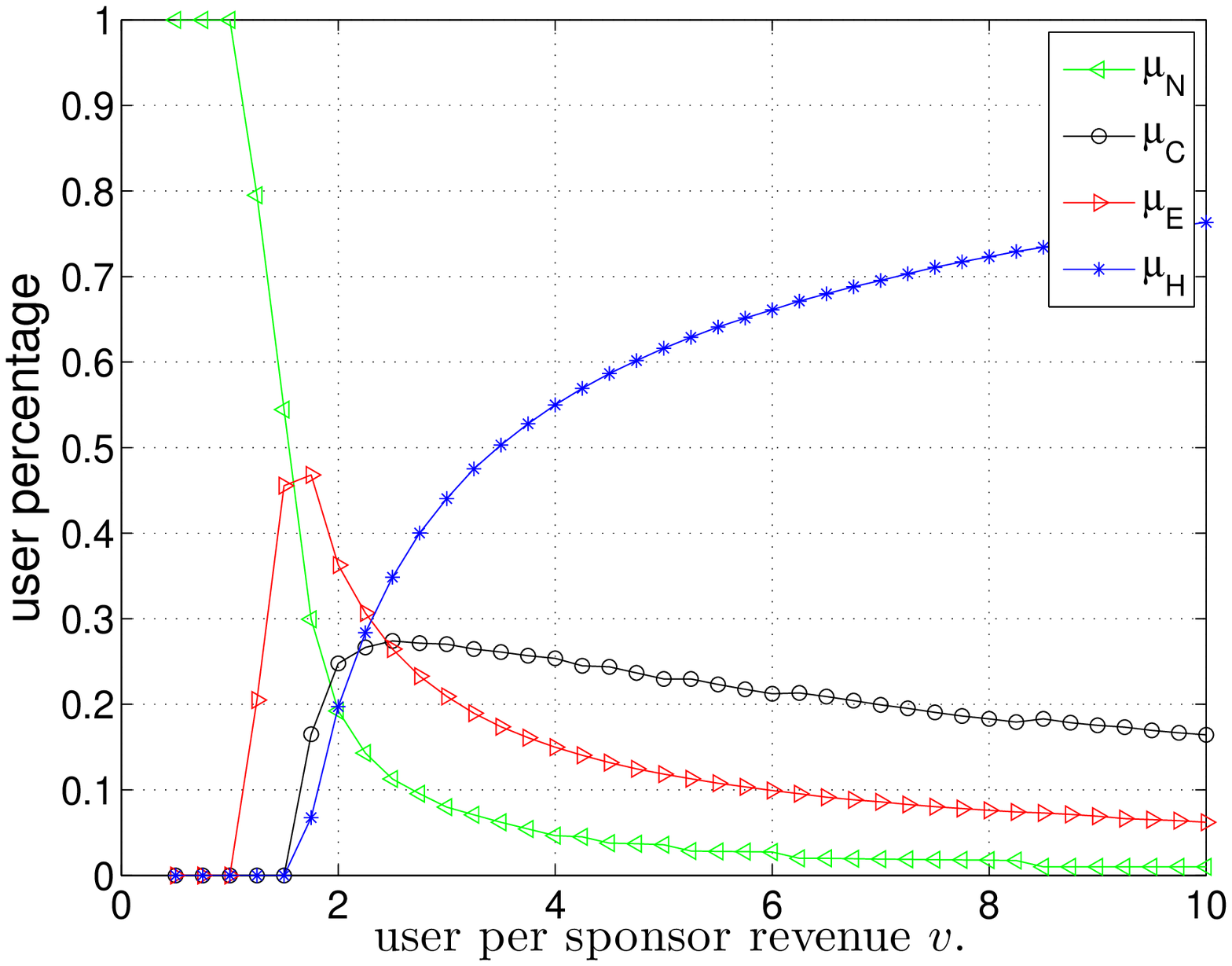}
\caption{Discuss user membership percentage with sponsor revenue $v$.}\label{fig:v_mu}	
  \end{minipage}
  \begin{minipage}[h]{.18\linewidth}
\includegraphics[width=0.9\textwidth]{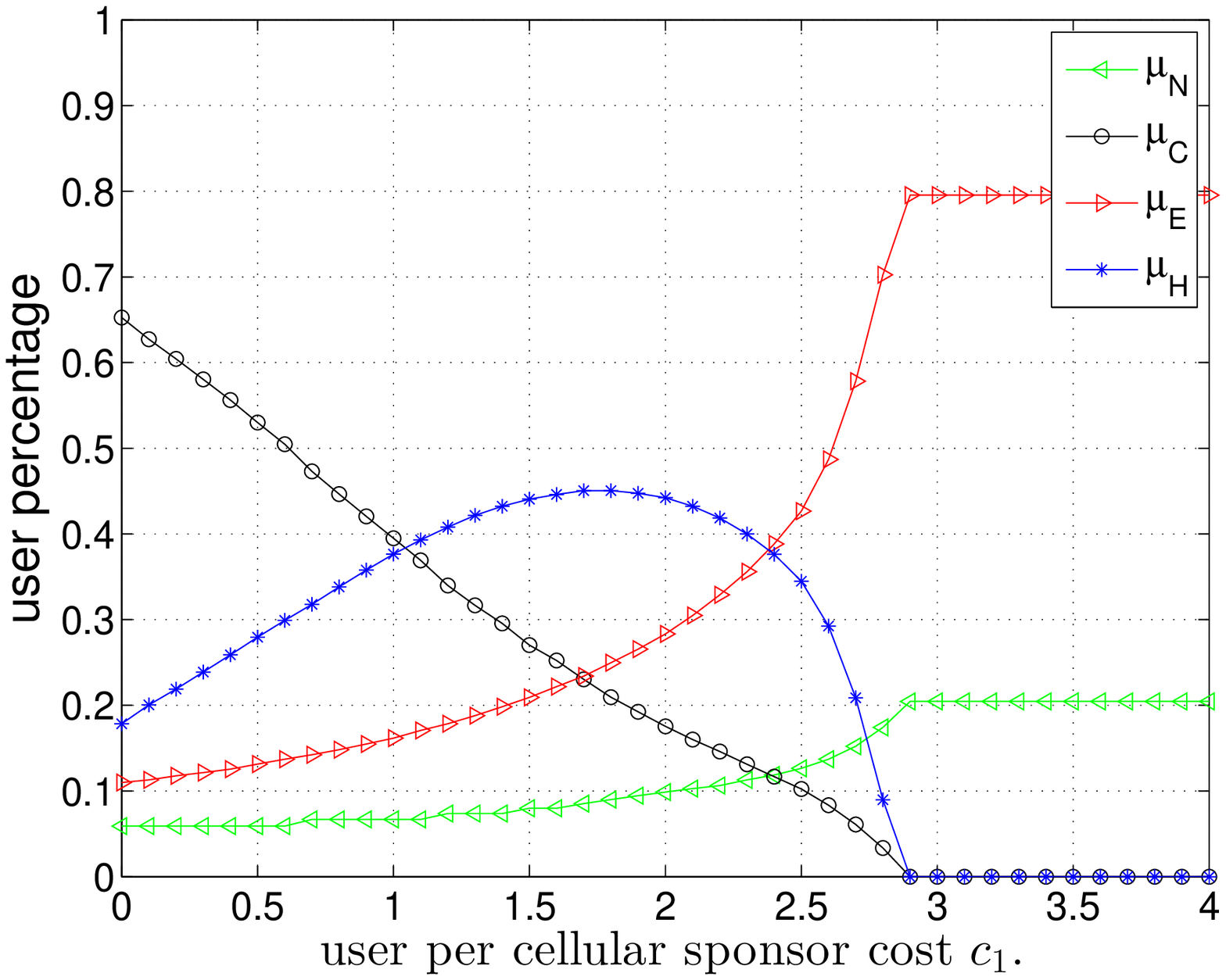}
\caption{Discuss user membership percentage with cellular sponsor cost $c_1$.}\label{fig:c2_mu}	
  \end{minipage}
  \begin{minipage}[h]{.18\linewidth}
	\includegraphics[width=1.08\textwidth]{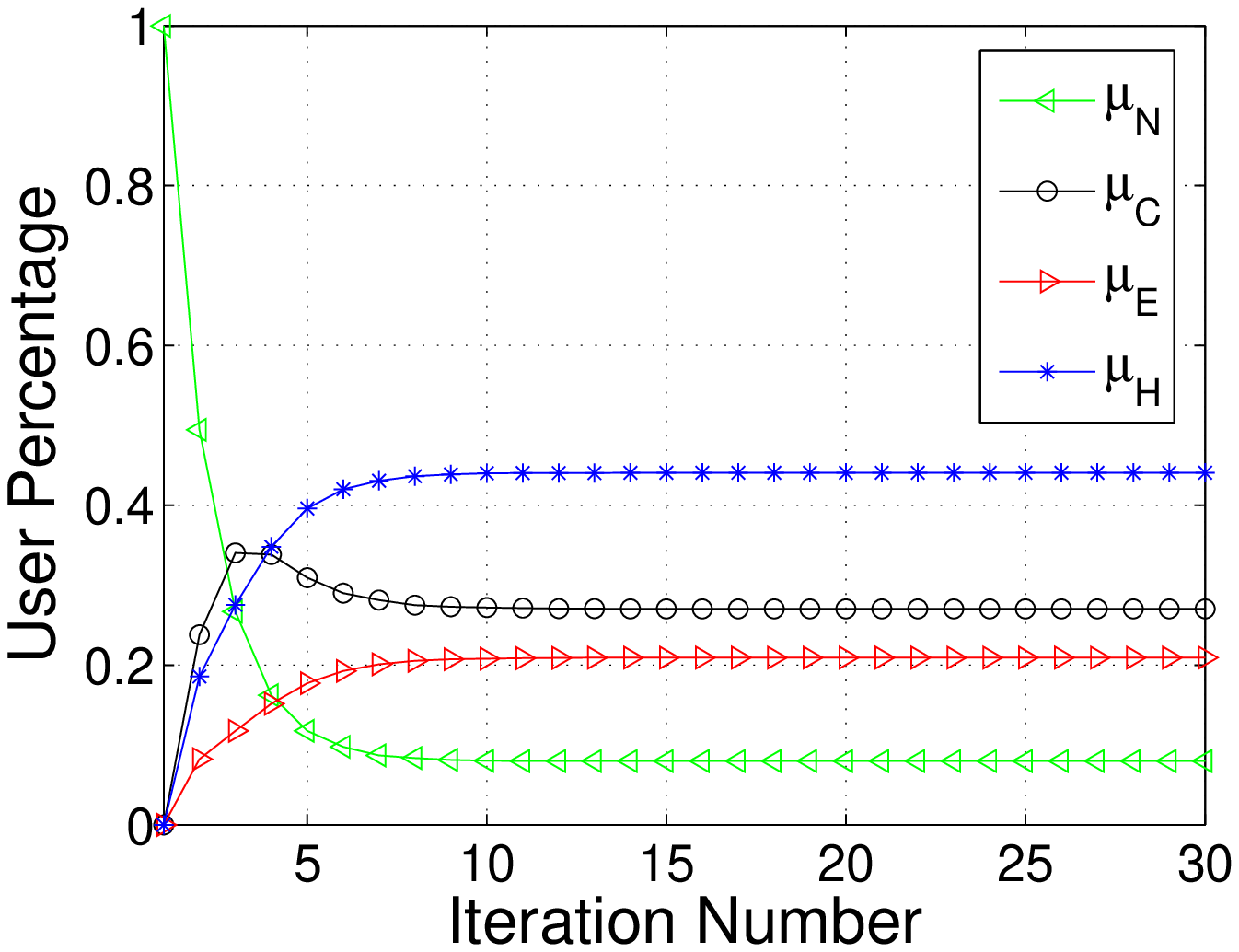}
	 \caption{Dynamic Selection in Stage II.}\label{fig:dynamic2}	
  \end{minipage}
  \begin{minipage}[h]{.18\linewidth}
     \includegraphics[width=1\textwidth]{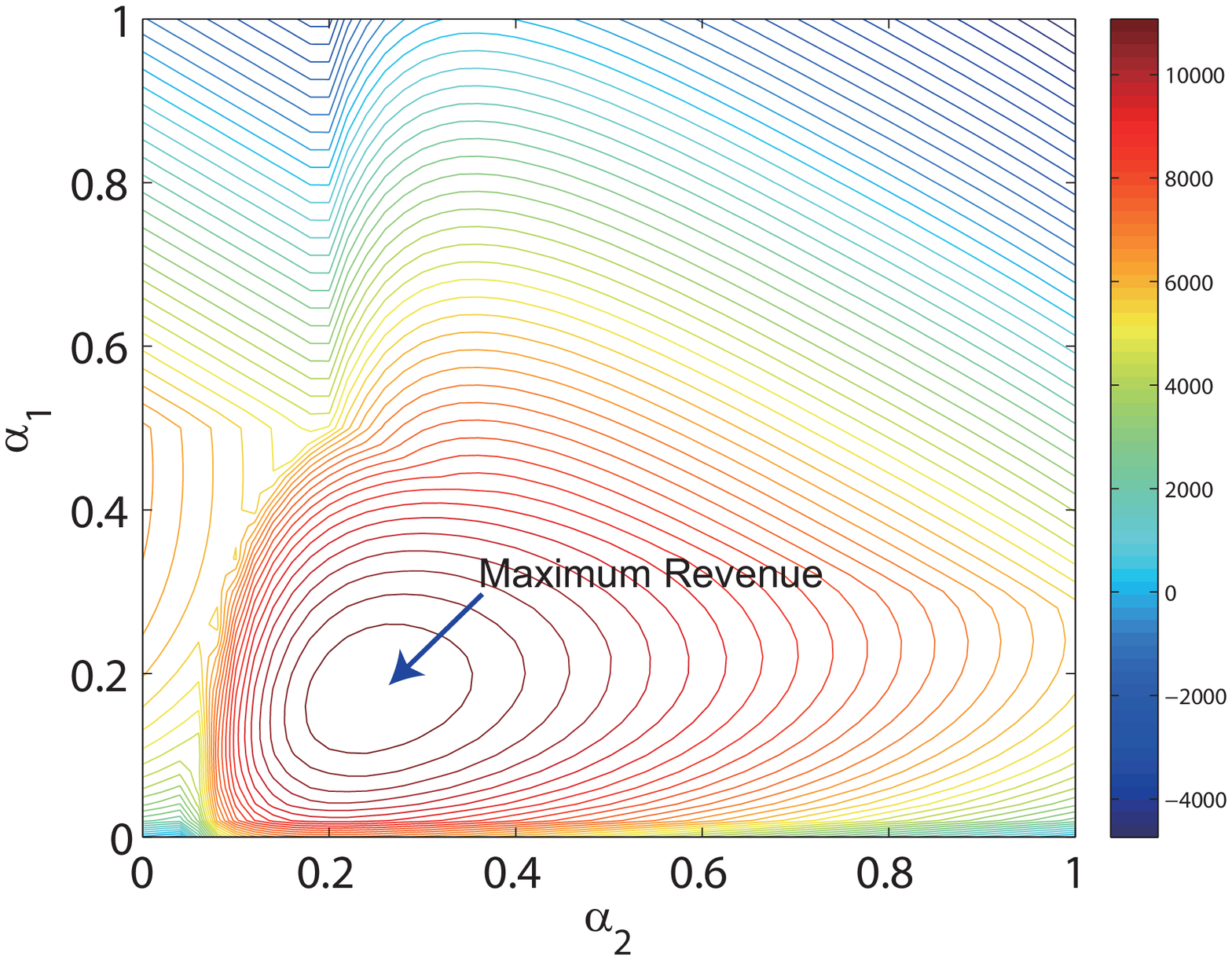}
	 \caption{The CP's Revenue v.s. Budgets.}\label{fig:optimal1}
  \end{minipage}
    \begin{minipage}[h]{.18\linewidth}	
     	 \includegraphics[width=1.08\textwidth]{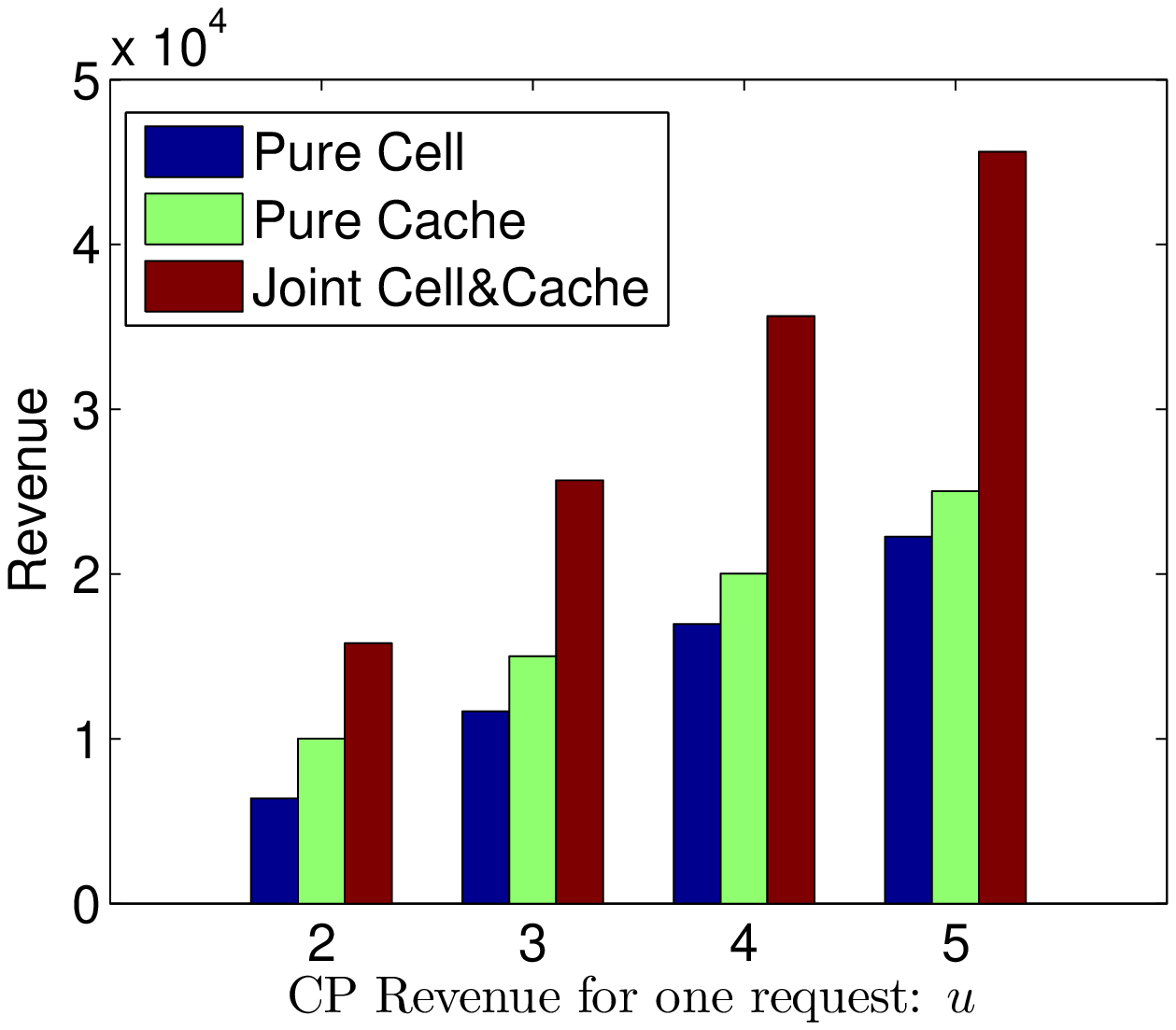}
	 \caption{The CP's Revenue v.s. Single Request Revenue.}\label{fig:revenue_pure_joint}	
  \end{minipage}
  \vspace{-5mm}
  \end{figure*}
We perform numerical studies in a mobile network, and verify the equilibrium analysis. We simulate a system in which there are 10,000 mobile video users, and choose system parameters as follows: $v=3$, $c_1=1.5$, $c_2=1$, $\phi=0.1$, $u = 3$, $h_1=1.5$, $h_2=2$, $S=1000$, $\alpha_1=2000$, and $\alpha_2=2000$.

\subsection{Discussion on Parameters}


We first discuss the impact of user sponsor revenue on users' sponsor membership decision. The user sponsor revenue is measured by the per sponsor revenue $v$. Fig.~\ref{fig:v_mu} demonstrates the user membership percentages under different user sponsor revenue $v$. We can see from Fig.~\ref{fig:v_mu} that the percentage of \textbf{HybridSp} users is monotone increasing with $v$, because the sponsorship becomes more beneficial to users, which attracts users to join both sponsorships. We notice that $\mu_N$ is monotone decreasing with $v$. From the CP's perspective, in order to attract more users to utilize the sponsorship schemes, the CP should know which contents are most valuable to users and sponsor the users with free request to the contents.

Then, we discuss the parameter $c_1$, \emph{i.e.,} the per sponsor cost in \textbf{CellSp}. Fig.~\ref{fig:c2_mu} show the user membership percentages under different cellular sponsor cost $c_1$. We can observe from Fig.~\ref{fig:c2_mu} that the percentage of users selecting \textbf{CellSp} decreases and that of users selecting \textbf{EdgeSp} increases with $c_1$. This is because users need to bear higher cost in \textbf{CellSp}, and hence some of them switch to \textbf{EdgeSp} with relatively lower cost. Note that when $c_1\ge 3$, no user will choose \textbf{CellSp}.


\subsection{Dynamic User Selection and Algorithm in Stage II}

Fig.~\ref{fig:dynamic2} shows the process of dynamic member selection in Stage II. Specifically, it illustrates the dynamics of membership percentages $\mu_N$, $\mu_C$, $\mu_E$, and $\mu_H$. In this example, the membership dynamics converge to the equilibrium within about $10$ iterations, the membership percentage is $\mu_N=0.09,\; \mu_C=0.28, \;\mu_E=0.21, \;\mu_H=0.42$.

\subsection{CP's Decision and Revenue in Stage I}

We now discuss how the CP can optimizes its budgets. Specifically, the CP can determine the proper $\alpha_1$ and $\alpha_2$ based on the users' equilibrium sponsorship selections.

Fig.~\ref{fig:optimal1} shows the contours of CP revenue with respect to $\alpha_1$ and $\alpha_2$. In this case, the optimal budgets are $\alpha_1^*=0.2$ and $\alpha_2^*=0.26$. Furthermore, we investigate the CP's revenue under pure cellular sponsoring, pure cache sponsoring, and joint cellular and caching sponsoring in Fig.~\ref{fig:revenue_pure_joint}. With proper budget setting, the joint sponsor schemes can increase its maximum revenue up to $105\%$ and $85\%$, respectively, comparing with cases with pure cellular sponsoring and pure edge cache sponsoring.

\section{Conclusion}


In this work, we studied how the edge caching will affect the CP's data sponsoring strategy as well as the users' behaviors and the data market.
Specifically, we considered a single CP who offers both the edge caching service and the data sponsoring service to heterogeneous mobile video users.
We formulated the interactions of the CP and the users as a two-stage Stackelberg game, with the CP as leader and the users as followers, and analyzed the sub-game perfect equilibrium (SPE) systematically.
Numerical results showed that by introducing the edge caching, the CP can increase his revenue up to $105\%$.

\end{document}